\documentclass[a4paper,12pt,epsf,amssymb,citesort]{article}

\usepackage{tabularx}
\usepackage{array}
\usepackage{graphics}
\usepackage{graphicx}
\usepackage{epsfig}
\usepackage{amsmath}
\usepackage{amssymb}
\usepackage{color}
\usepackage{cite}
\usepackage{verbatim}
\newcommand{\pslash}{p \!\!\!/}

\begin{document}

\date{\today}

\title{\bf $\mu^2$ dependent deviation of the non perturbative $Z_A^{MOM}$ from the true axial renormalisation constant, implied by Ward identity}
\author{Ph.~Boucaud$^a$,
J.P.~Leroy$^a$, A.~Le~Yaouanc$^a$, J. Micheli$^a$,\\ O.~P\`ene$^a$, J.~Rodr\'{\i}guez--Quintero$^b$}

\maketitle

\vspace{-10.5cm}
\begin{flushright}
\begin{tabular}{l}
{\tt LPT-Orsay-15-26}\\{\tt UHU}\\
\end{tabular}
\end{flushright}
\vspace{10.5cm}

\begin{center}
$^a$Laboratoire de Physique Th\'eorique ; Univ. Paris-Sud ; CNRS; UMR8627 ;\\
 B\^atiment 210, Facult\'e des Sciences,  91405 Orsay Cedex.\\
$^b$ Dpto. F\'isica Aplicada, Fac. Ciencias Experimentales,\\
Universidad de Huelva, 21071 Huelva, Spain.
\end{center}

\begin{abstract}
  It is recalled why, as already stated in a previous paper, there seems to be an inconsistency in identifying the  {\bf non perturbative} $Z_A^{MOM}$ as the renormalisation of the axial current, or equivalently, in setting as normalisation condition that the renormalised $vertex=1$ at $p^2=\mu^2$ at some renormalisation scale $\mu$, where $p$ is the momentum in the legs. Indeed, unlike the vector case, the Ward-Takahashi (WT) identity for the axial current is shown to imply both the renormalisation scale independence of $Z_A$ and a $\mu^2$ dependence of $Z_A^{MOM}$. This $\mu^2$ dependence is simply related to certain invariants in the pseudoscalar vertex and can persist in the chiral limit due to the spontaneous breaking of chiral symmetry (pion pole). It is seen clearly in the $\mu^2$ dependence of some lattice calculations of $Z_A^{MOM}/Z_V^{MOM}$ near the chiral limit.
\end{abstract}

\section{Introduction}
The non perturbative MOM renormalisation scheme for lattice as introduced in Martinelli et al. Nucl.Phys. B445 (1995) 81-108, \cite{martinelli9411010}, and inspired by the corresponding continuum scheme of Georgi and Politzer \cite{politzer},  has represented an imprtant progress in lattice QCD. It is very intuitive and easy to handle by pedestrians because it relies on the vertices and propagator functions which are familiar in pertrubative QCD. This is to be compared with later schemes like the one of Alpha \cite{luscher9611015}, which is rigourous, but requires much more effort to understand and many technicalities.

Nonetheless, it has features coming from its direct hadronic meaning that precisely complicate the matching with the usual perturbative schemes which is the final goal. It has a non logarithmic, power dependence on $\mu^2$ generated by OPE power corrections \cite{BoucaudPRD} and particle poles (although the latter are lying outside the Euclidean range). It may have critical chiral behaviour unlike the Alpha prescription, as has been underlined some time ago for the pseudoscalar $Z_P$ \cite{cudell}, leading to the recipe of "extracting the pion pole".

Moreover, it seems  to have often escaped the attention that there is also in principle {\bf an inconsistency in introducing $Z_A^{MOM}$ as the renormalisation of the axial current}, as stated in our paper PhysRevD.81.094504 \cite{massfunction}. This manifests itself in a $\mu^2$ dependence of $Z_A^{MOM}/Z_V^{MOM}$ which persists in the chiral limit, even with an explicitly chiral invariant action, due to the spontaneous breaking of the symmetry, see our earlier paper\cite{quarkprop}, section 8.1..

Since this statement was presented in a rather long paper devoted to several topics, we think useful to recall the arguments in a clearer and more explicit manner, and in the same course to correct some sloppy notations of the paper.

It must also be said that having reread the basic paper \cite{martinelli9411010}, we have rediscovered that there was a discussion in it having connection with the present one, although the conclusion seems different : we extract a finite effect in the chiral limit, which does not appear in their approach. The reason will appear after having presented our own discussion, in a separate section \ref{martinelli}. It requires an  examination of the interplay of the chiral and $q \to 0$ limits.

\section{Practical meaning of the problem}

One must warn from the beginning about the practical meaning of the problem . The problem is found to disappear at large $\mu^2$, so that one may claim that it is not real since anyway, the non perturbative MOM scheme is meant precisely to be applied at such large $\mu^2$. More precisely, for $Z_{V,A}$ one seems to be free to choose any $\mu^2$, therefore, it would suffice to work at such a large $\mu^2$  with $Z_A^{MOM}$. In fact the effect seems anyway to be small already about $\mu=2~$GeV or beyond, see fig. 5  in the quoted paper \cite{quarkprop}, which is now the commonly adopted value for non perturbative renormalisation. Therefore, the effect is perhaps not worrying practically, in contrast to the pion pole in $Z_P^{MOM}$. 

However it is worth in general being clear on theoretical principles. But also working at large $\mu^2$ presents well-known practical problems.

1) one may always wonder how large $\mu^2$ must be. 

2)  measuring Green functions at large momenta requires the extraction of large artefacts, and this was the initial motivation for non perturbative renormalisation : to avoid working at too large momenta. 

On the other hand, if $\mu^2$ is not sufficiently large, one has to ``extract" a physical $\mu^2$ dependence, while artefacts may still be non negligible (one cannot exclude non canonical artefacts at small $\mu^2$).

For all these reasons, it is useful to know about the possible causes of momentum dependence, either ``physical" as the present one (physical with many quotation marks), or artefactic.
 
One must add that at present little effort has been devoted to determine the actual magnitude of the effect (see the end of the text).

\section{Definitions and generalities}

Let us first fix the notations that we will use.
We will use all along the Euclidean metrics.  The continuum quark propagator
is a $12 \times 12$ matrix $S(p_\mu)$ for 3-color and 4-spinor indices. One can take into account Lorentz (in fact $O(4)$) invariance and discrete symmetries, as well as color neutrality of the vacuum by expanding the inverse propagator according to :
\begin{eqnarray}\label{Sm1}
S^{-1}(p) = \delta_{a,b} Z_\psi(p^2) \left( i\,\pslash + M(p^2)\right)
\end{eqnarray}
where $a,b$ are the color indices. ``$Z_\psi(p^2)$" is a standard lattice notation, alluding to the role it plays as a renormalisation constant for the quark field in the standard Georgi-Politzer MOM renormalisation, where  $Z_2^{MOM}(\mu^2)=Z_\psi(\mu^2)^{-1}$. But let us stress that here it is not by itself a renormalisation constant. On the other hand, $M(p^2)$ is the mass function, which is one possible concept of mass, introduced by Georgi and Politzer. $M(p^2)$ is {\bf UV finite}, since it is the ratio of two quantities renormalised by the same factor $Z_2$. It is identical with the MOM renormalised mass at scale $\mu^2=p^2$, see also below.

Let us consider a colorless local two quark operator $\bar q {\cal O} q$.
 The corresponding three point Green function  $G$ is defined by
\begin{eqnarray}\label{G}
G(p, q) = \int d^4x d^4y~e^{i p \cdot y + i q \cdot x} 
< q(y)\bar q(x) {\cal O} q(x) \bar q(0) >
\end{eqnarray}
It is a $4 \times 4$ matrix in Dirac space.
The associated vertex function is then defined by amputation
of quark propagators on both sides :
\begin{eqnarray}\label{gammamu1}
\Gamma(p, q) = S^{-1}(p) \,G(p, q)\,S^{-1}(p+q)
\end{eqnarray}
In the note, we will often restrict ourselves to the case where
the operator carries a vanishing momentum transfer $q_\mu=0$. We will then omit to write $q_\mu=0$ and we will moreover understand $\Gamma(p)$  without $R$ index as the {\bf bare} vertex function (computed on the lattice).

Now, Lorentz covariance and discrete symmetries allow to write for the axial vertex $\Gamma_{A \mu}(p,q)$ at $q=0$:
\begin{eqnarray}\label{gammamu2}
\Gamma_{A \mu}(p) = \delta_{a,b} [g_A^{(1)}(p^2) \gamma_\mu  \gamma_5+
 i g_A^{(2)}(p^2) p_\mu  \gamma_5 + \nonumber \\
g_A^{(3)}(p^2) p_\mu \pslash  \gamma_5+ ig_A^{(4)}(p^2) [\gamma_\mu,\pslash] \gamma_5 ]
\end{eqnarray}
which should be obeyed approximately on the lattice, as we checked.

On the other hand, we need the pseudoscalar vertex at $q \neq 0$ :
\begin{eqnarray}\label{gamma5prime}
\Gamma_5 (p,q) = \delta_{a,b} \left[ g_5^{(1)}(p,q) \gamma_5 + i g_5^{(2)}(p,q) \gamma_5 ( p_\mu \gamma_\mu ) + \right . \nonumber \\ 
 \left. i g_5^{(3)}(p,q)\gamma_5 \gamma_{\mu} q^{\mu}+ g_5^{(4)}(p,q) \gamma_5 [\gamma_{\mu} q^{\mu},\pslash] \right]
\end{eqnarray}
where the quark momenta are $p,p+q$ and the $g_5^{(i)}(p,q)$'s are invariant functions of the momenta alone. For brevity, the first two invariants are denoted by the same symbol at $q=0$, i.e. :
\begin{eqnarray}
g_5^{(1,2)}(p^2)=g_5^{(1,2)}(p,q=0)
\end{eqnarray}
More explicitly, the dependence of $g_5^{(i)}(p,q)$ in $p,q$ is :
\begin{eqnarray}
g_5^{(i)}(p,q)=g_5^{(i)}(p^2,q^2,p.q)
\end{eqnarray}

In the following, the color factors $\delta_{a,b}$ will be skipped.
\vskip0.2cm

\subsection{Renormalisation in general}
Without requiring any specific renormalisation scheme, we have to refer to the renormalisation, because the Ward-Takahashi(W-T) identities should be imposed on the renormalised theory, and not on the bare quantities (we do not consider anomalies). The corresponding renormalised quantities are denoted by a sub- or superindex  ${\rm R}$. We then draw the consequences for the specific MOM scheme.

$Z_2$ denotes as usual the fermion field or propagator renormalisation according to :
\begin{eqnarray}
q=\sqrt{Z_2} q_R \nonumber \\
S(p)= Z_2 S_{\rm R}(p)
\end{eqnarray}
Let us recall that the corresponding renormalised vertex functions are defined through:
\begin{eqnarray}
\Gamma(p)=Z_2^{-1} Z_{\cal O}^{-1} \Gamma_{\rm R}(p) , 
\end{eqnarray}
where the necessary subindices are implicit for each type of vertex; $Z_{\cal O}$ is the renormalisation of the composite operator, namely a current or density operator : ${\cal O}=j_V,j_A,P_5$ ; the $Z_2$ factor is to take into account the amputation~\footnote{Note that the standard definition of renormalisation constants is to \underline{divide} the bare quantity by the renormalisation constant to obtain the renormalised quantity (except for photon or gluon vertex renormalisation factors $Z_1$ which we do not use). In principle, renormalisation of composite operators, for instance $Z_V$, should be defined similarly. We have followed this convention in our works on gluon fields, for the renormalisation of $A^2$. But, in the case of \underline{quark} composite operators,
an opposite convention has become standard in lattice calculations : $(\bar q {\cal O} q)_{bare}=Z_{\cal O}^{-1} (\bar q {\cal O} q)_R$ ; we feel compelled to maintain this convention for the sake of comparison with parallel works on the lattice. This explains our writing of the renormalised vertex function.}.

The lattice calculations, being done at a finite cut-off,  generate, as other regularisation schemes,  finite ${\cal O}(g^2)$ effects, due to additional divergencies multiplying the $a$ terms (which have higher dimension), which vanish slowly with the inverse cutoff or lattice unit $a$, and are included in the factors $Z_2,Z_V,~Z_A$. There are also terms with powers of $a$ which we do not write. $Z_V,~Z_A$ are independent of the renormalisation scheme up to such terms. The fact that we do not include such terms means that our equations should hold only sufficiently close to the continuum.

\subsection{The axial Ward identity}

Let us develop the consequences of the axial W-T identity. We define a bare mass $\rho$ through the equation 
\begin{eqnarray} \label{bareAx}
\partial_{\mu} (j_A)^{\mu}= 2 \rho P_5 
\end{eqnarray}
(the notation $\rho$ is old and unsuggestive of a mass, but it avoids any ambiguity in a world where there are so many masses). In renormalised form :
\begin{eqnarray}
\partial_{\mu} (j_A)_{\rm R}^{\mu}= 2 m_R  (P_5)_R 
\end{eqnarray}
$m_{\rm R}$ is the renormalised mass in the considered scheme, which then satifies the relation :
\begin{eqnarray}\label{mR}
m_{\rm R}=Z_P^{-1} Z_A \rho.
\end{eqnarray}
To exploit fully the identities, one  has to return first to the general case $q=p'-p \neq 0$.
Since they reflect the symmetries of the physical theory, the naive Ward identities should a priori hold for the renormalised Green functions (except for anomalies) and at infinite cutoff, which means : 
\begin{eqnarray}\label{AwardR}
q_{\mu}\ (\Gamma_A)_{\rm R}^{\mu}(p,q)=-i\,(S_{\rm R}^{-1}(p+q)\gamma_5 +\gamma_5 S_{\rm R}^{-1}(p))+i \,2 m_{\rm R} (\Gamma_5)_{\rm R}(p,q)
\end{eqnarray}
Returning then to bare quantities which are the ones actually measured on the lattice, one gets, 
multiplying both sides by  $Z_2^{-1}$ and using $m_{\rm R}=Z_P^{-1} Z_A \rho$ :
\begin{eqnarray}\label{AwardB}
Z_A~q^{\mu}(\Gamma_A)_{\mu}(p,q) = -i\, (S^{-1}(p+q)\gamma_5+\gamma_5 S^{-1}(p))+i\,2 Z_A \rho \Gamma_5(p,q),\quad
\end{eqnarray}
which depends only on one renormalisation constant $Z_A$, and bare, renormalisation scheme independent, quantities. 

Since eqn. (\ref{AwardB} ) has been established  without any specification of the renormalisation scheme, it shows that $Z_A$ also is {\bf independent of the renormalisation scheme}. Therefore, one should expect $Z_A^{MOM}$ to be equal to $Z_A$. But {\bf this is not the case}, as the same identity eqn. (\ref{AwardB}) shows, see the demonstration below.

\subsection{MOM scheme}

Let us now introduce the MOM scheme. It must be first defined for the propagator, through conditions 
at some normalisation momentum
$p^2=\mu^2$, originally due to Georgi and Politzer \cite{politzer}:
\begin{eqnarray}
S_R^{-1}(\mu) = \delta_{a,b} \left( i\,\pslash + m_{\rm R}^{MOM}\right) \vert_{p^2=\mu^2},
\end{eqnarray} 
which means 
\begin{eqnarray}
Z_2^{MOM}=Z_{\psi}(\mu^2)^{-1} 
\end{eqnarray}
according to eqn. (\ref{Sm1}). Also, it means that the renormalised mass is then 
\begin{eqnarray} \label{MOMmass}
m_{\rm R}^{MOM}=M(\mu^2),
\end{eqnarray} 
 i.e. it is the mass function at $p^2=\mu^2$.

As to quark current vertices,  it is then commonly accepted that they can be renormalised analogously by setting $(g_{V,A,5}^{(1)})_{\rm R}(p^2=\mu^2)=1$ \footnote{We set standard conditions on one invariant. We are aware that others may be set by combining several invariants in a trace. They lead to complications in the discussion.}. This leads to the well-known "renormalisation constants", for example :
\begin{eqnarray}
``Z_{V,A}^{MOM}"=Z_{\psi} (\mu^2)/g_{V,A}^{(1)}(p^2=\mu^2)
\end{eqnarray}
The quotation marks are provocative and aim at signalling that they may not be the true renormalisation constants. Indeed, it must be stressed that having fixed the renormalisation of the propagator, {\bf the renormalisation conditions of vertices cannot be imposed freely} : they must be constrained by the renormalised WT identities (similarly to the Slavnov identities for QCD vertices); one is then not allowed to set $(g_{V,A,~5}^{(1)})_{\rm R}(p^2=\mu^2)=1$ freely.

\section{An equation for $``Z_A^{MOM}"/Z_A$ by derivation of the WT identity near $q^{\mu}=0$}
 
Let us first recall the very simple argument concerning the vector case. In the vector case, the Ward identity is very simple : 
\begin{eqnarray}\label{Vward4}
Z_V~q^{\mu}(\Gamma_V)_{\mu}(p,q) = -i\, (S^{-1}((p+q)-S^{-1}(p)) .
\end{eqnarray}
Then, as is well known, by derivation with respect to $q^{\mu}$, one gets among other relations:
\begin{eqnarray} \label{Vward5}
Z_V=Z_{\psi}(p^2)/g_V^{(1)}(p^2)
\end{eqnarray}
where $g_V^{(1)}(p^2)$ is the coefficient of the $\gamma_{\mu}$ term
in the Lorentz decomposition of the vector vertex. $Z_V$ has thus be determined independently of any choice of renormalisation scheme : it is indeed independent of the scheme, being expressed in terms of bare quantities.

It must be  noticed, of course, that the r.h.s. of eqn. (\ref{Vward5}) is nothing else than the MOM renormalisation constant at $\mu^2=p^2$ defined by the renormalisation condition $(g_V^{(1)}(\mu^2))_R=1$ , i.e. :
\begin{eqnarray} \label{Vward6}
Z_V=``Z_V^{MOM}(\mu^2)"
\end{eqnarray}
Therefore, ``$Z_V^{MOM}(\mu^2)$" is indeed the expected renormalisation of the vector current and we may abandon the quotation marks \footnote{It must be observed that we stick strictly to definition of MOM condition through one invariant $i=1$. We do not consider sums over several invariants as done sometimes}.

Therefore also, up to now, everything is well with W-T identities in the MOM non perturbative scheme.

Now, for the axial case, comes the inconsistency . As in the vector current case, the axial W-T identity will give a constraint on the axial vertex at $q=0$ by taking the derivative of eqn. \ref{AwardB} with respect to $q$ at $q=0$. We get :
\begin{eqnarray}\label{Award4}
Z_A (\Gamma_A)_{\mu}= -i  \frac {\partial}{\partial p^{\mu}} S^{-1}(p) \gamma_5 + 2 i\, Z_A \rho \frac {\partial}{\partial q^{\mu}} \Gamma_5(p,q)
\end{eqnarray}

It must be stressed that not only this relation (\ref{Award4}) is more complex than in the vector case  (\ref{Vward4}) , due to the $\Gamma_5(p,q)$ contribution, but also the latter does not vanish in general even in the chiral limit, because of the pion pole in $\Gamma_5(p,q)$ \footnote{A somewhat different expression was given in the previous paper, due to a confusion with an older definition of $Z_A^{MOM}$ through traces.}. This is one more manifestation of the spontaneous breaking of chiral symmetry. Also, of course, it must be recalled that, on the lattice, the Ward identity
is not exact, but holds only up to artefacts, because we work at finite cutoff, and the deviation could be found very large in some cases. 

Let us comment more on the chiral limit. The demonstration has been done away from the chiral limit, and the relation is at $q^{\mu}=0$. Now, the chiral limit of the r.h.s. of eqn. (\ref{Award4}) is regular since the coupling of a pseudoscalar pion to the axial current has a factor $q_{\mu}$ :
\begin{eqnarray}
\langle \pi\vert (j_A)_{\mu}\vert 0 \rangle \propto f_{\pi} q_{\mu} .
\end{eqnarray}
and vanishes at $q^{\mu}=0$. Therefore, the limit of the r.h.s. must be also regular, although non zero. This will be shown explicitly below.

From the equation (\ref{Award4}), one deduces that $Z_A \neq ``Z_A^{MOM}"$, where $``Z_A^{MOM}"$ is defined, in parallel with $Z_V^{MOM}$, as $Z_{\psi}(p^2=\mu^2)/g_A^{(1)}(p^2=\mu^2)$. This is due to the derivative of the pseudoscalar term. In fact $Z_{\psi}(p^2=\mu^2)/g_A^{(1)}(p^2=\mu^2)$ is not even independent of $\mu^2$; one can hope only that it reaches $Z_A$ at large $\mu$ ; then, it would be perhaps better to discard this MOM definition, since the word is misleading. 

Let us show this statement in more detail, by tracking the contributions having the $\gamma_{\mu} \gamma_5$ structure of the undressed vertex, to match the invariant $g_A^{(1)}(p^2=\mu^2)$ in the axial vertex. From now on, we keep to the general $p$ instead of setting $p^2=\mu^2$, which is not useful. In $\Gamma_5(p,q)$, written in full in equation (\ref{gamma5prime}) the relevant term is obviously
\begin{eqnarray}
\Gamma_5(p,q)=...+i g_5^{(3)} \gamma_5 \gamma_{\mu} q^{\mu} +...
\end{eqnarray}
The derivative at $q=0$ gives a contribution $2 i Z_A \rho \times -i g_5^{(3)}(p^2) \gamma_{\mu} \gamma_5$ to eqn. (\ref{Award4}) \footnote{One must be aware that since there are other possible definitions of $Z_A^{MOM}$, involving traces, one would obtain for them a similar equation, but with a contribution of different invariants of the pseudoscalar vertex, for instance the derivative of $g_5^{(2)}(p,q)$ for a trace on $\gamma_{\mu} \gamma_5$.} . Therefore :
\begin{eqnarray}
Z_A g_A^{(1)}(p^2)=Z_{\psi}(p^2)+2 Z_A \rho g_5^{(3)}(p^2)
\end{eqnarray}

or 
\begin{eqnarray} \label{discrepancy}
Z_A /Z_A^{MOM}(p^2)=1+2 Z_A \rho~g_5^{(3)}(p^2)/Z_{\psi}(p^2)
\end{eqnarray}
The derivation on the propagator coefficients $Z_{\psi}(p^2)$ and $Z_{\psi}(p^2) M(p^2)$, as well as on the other invariants in $\Gamma_5$ is seen to give contributions to the other invariants in the expansion of the axial vertex, eqn. (\ref{gammamu2}).

One can express the result (\ref{discrepancy}) in a more striking form, returning to the renormalised axial vertex, and to $\mu^2$ as the MOM renormalisation scale for the propagator. Since $Z_2(\mu^2)=1/Z_{\psi}(\mu^2)$, $Z_A /Z_A^{MOM}(\mu^2)$ is nothing else than $Z_A Z_2 g_A^{(1)}(\mu^2)=(g_A^{(1)})_R~(p^2=\mu^2)$. Then the relation is nothing but :
\begin{eqnarray} \label{discrepancy2}
(g_A^{(1)})_R(p^2=\mu^2)=1+2 Z_A Z_2(\mu^2)\rho~g_5^{(3)}(\mu^2)
\end{eqnarray}
which exhibits clearly the statement that $(g_A^{(1)})_R(p^2=\mu^2)$ cannot be chosen arbitrarily 
once the propagator has been renormalised : its value is completely determined, and in particular {\bf it cannot be set to $1$}.

We see no reason why $g_5^{(3)}(p^2)$ should vanish. Rather, it is clear that it contains a pion pole contribution, since $\gamma_{\mu} q^{\mu} \gamma_5$ is a known structure in the Bethe-Salpeter vertex function of the pion, see for instance the appendix of Nambu and Jona-Lasinio, Phys.Rev. 124 (1961) 246-254 \cite{nambu}.  It is reasonable to suppose that the effect vanishes by powers at large $p^2$, since this is the general behaviour of Green functions. One must note however that the transition to $0$ as seen from Fig. 5 of \cite{quarkprop} is rather abrupt. 

The chiral limit of the deviation is finite but non zero as announced, since the pole $1/m_{\pi}^2$ of $g_5^{(3)}(p^2)$ is multiplied by the factor $m_q$ \footnote{There has been statements that $Z_A^{MOM}/Z_V^{MOM}=1$ in the chiral limit, but they rest on the assumption that the vacuum is chiral symmetric, so they are valid only asymptotically in $p^2$}.

We have not heard of such an inconsistency of the $MOM$ definition of the axial vertex renormalisation in perturbative QCD.  At least the deviation should vanish in the chiral limit in perturbative QCD, since there would no longer be a pion pole to compensate the $m_q$ factor. On the other hand, it remains to be known by explicit calculation what happens at $m_q \neq 0$.

Finally, let us note that if one were using definitions of $Z_A^{MOM}$ with traces, as has been usual for some time, one would have to include other invariants of $\Gamma_5$ in the expression of $Z_A/Z_A^{MOM}$.

\subsection{Connected observation of momentum dependence of $Z_A^{MOM}/Z_V^{MOM}$ in lattice calculations} 

The effect may be exhibited most clearly with chiral invariant actions at $m_q=0$ where $Z_A/Z_V=1$ is expected to hold exactly. And indeed, it seems to have been seen in certain lattice simulations of $Z_A^{MOM}/Z_V^{MOM}$ with chiral symmetry preserving actions, showing near the chiral limit a decrease from $1$ with decreasing $q^2$ (Dawson, with domain wall fermions \cite{dawson}  ; our paper Phys.Rev.D74:034505,2006 \cite{quarkprop} on the quark propagator with Ginsparg-Wilson action, especially fig. 5), while it reaches $1$ at large momentum. But in neither of these works, was it possible to separate cleanly the effect from artefacts, and the works should be redone. In connection, a lattice calculation of the new invariants in the pseudocalar vertex should be made to ascertain the estimate obtained in eqn. (\ref{discrepancy}).

\section{Relation with the discussion of Martinelli et al.} \label{martinelli}
In fact, there is in paper \cite{martinelli9411010} a discussion also concluding to a difference $Z_A^{MOM} \neq Z_A$. Apparently it has not led to further discussion. It has a connection with ours. Their argument also rests on the derivative of the W-T identity at  $q=0$, and they also conclude that the problem should disappear at large momenta. However, our finite result does not appear in their calculation.


\vspace {2cm}
Why ? This requires a rather long explanation. The first difference is that they work strictly at $m_q=0$, without taking a limit from $m_q \neq 0$. At first sight, this approach is quite opposite and could be expected to be incompatible with ours. Moreover, one could feel dangerous to work at $m_q=0$ and indeed we have preferred to start from $m \neq 0$. Nevertheless, we show finally that one could obtain the same result as in our approach for $m_q \to 0$. But this requires using the $m_q=0$ approach differently.

\vskip 0.2cm

1) Let us first compare the starting point of their discussion. 

Their argument is based on their eqn. (12) where the W-T identity {\bf has no pseudoscalar term in the r.h.s}. On the contrary, our pseudoscalar term in the r.h.s., is {\bf non zero even in the chiral limit (in the form $0/0$)}. This r.h.s. pseudoscalar term is crucial in our argument, and in fact the effect is present at any mass $m_q$, not only in the chiral limit. 

Moreover, in their treatment at $m_q=0$, they find by derivation of the W-T identity their eqn. (13), with some non explicited contribution to the derivative  of $q^{\mu}(\Gamma_A)_{\mu}(p,q)$ at $q^{\mu}=0$ due to the $q^2=0$ Goldstone pole in $(\Gamma_A)_{\mu}(p,q)$. 

We do not find such a contribution in our treatment with $m_q\neq 0$, because $(\Gamma_A)_{\mu}(p,q)$ is {\bf not singular at all at $q=0$}, the pole being shifted by $m_{ps}^2$. We obtain rather a $q_{\mu}/(q^2-m_{ps}^2)$, therefore a contribution $q^2/(q^2-m_{ps}^2)$ to $q^{\mu}(\Gamma_A)_{\mu}(p,q)$, whose derivative is $0$ :
 \begin{eqnarray}
\frac {\partial} {\partial q^{\mu}} q^2|_{q^{\mu}=0}=0
\end{eqnarray}
(We work here in Minkowski space-time for easiness)
\vskip 0.2cm

2) In our opinion, a $m_q=0$ approach, although perhaps virtually dangerous, is nevertheless possible with various precautions, as we show below\footnote{Let us recall that the mechanism of Nambu-Jona-Lasinio was partly illustrated  within this strict chiral symmetric situation, at $m_q=0$ \cite{nambu}}. But there is another problem in \cite{martinelli9411010}. One  decomposes $\frac {\partial} {\partial q^{\nu}} q^{\mu} (\Gamma_A)_{\mu}$ into a sum of two terms (l.h.s. of their eqn.(13):
\begin{eqnarray} \label{decomposition}
\frac {\partial} {\partial q^{\nu}} (q^{\mu} (\Gamma_A)_{\mu} )=\delta_{\nu}^{\mu}(\Gamma_A)_{\mu} +q^{\mu} \frac {\partial} {\partial q^{\nu}} (\Gamma_A)_{\mu} .
\end{eqnarray}

This decomposition has the drawback that both terms of the sum are singular at $q^2=0$, while their sum was regular : the singularities are $q_{\nu} 1/q^2$ in their first term, $- q_{\nu} 1/q^2$ in the second one. The finite difference is then not made explicit.  
\vskip 0.2cm

3) Let us now show that one can retrieve our result, at least in the chiral limit, through a calculation strictly done at $m_q=0$ but avoiding this decomposition.  

Let us consider the Goldstone contribution. There is no pseudoscalar term in the r.h.s. of W-T identity in the $m_q=0$ method, and the pole is now only present in $(\Gamma_A)_{\mu}$, and it has the form :
\begin{eqnarray}\label{Goldstone}
 \frac {q_{\mu}} {q^2} f_{\pi} \Gamma_{\pi} (p,q)
\end{eqnarray}
where $\Gamma_{\pi} (p,q)$ is the full pseudoscalar vertex of the pion  i.e. a $4 \times 4$ with a structure  parallel to the one of $\Gamma_5.$.

It gives a regular contribution to $q^{\mu} (\Gamma_A)_{\mu}$ : 
\begin{eqnarray}
q^{\mu} \frac {q_{\mu}} {q^2} f_{\pi} \Gamma_{\pi} (p,q)=f_{\pi} \Gamma_{\pi} (p,q)
\end{eqnarray}
The derivative of $q^{\mu} (\Gamma_A)_{\mu}$ at $q^{\mu}=0$ then has a term not present at $m_q \neq 0$ :
\begin{eqnarray}\label{goldstone}
f_{\pi} \frac {\partial} {\partial q^{\mu}}\Gamma_{\pi} (p,q)|_{q^{\mu}=0}
\end{eqnarray}

However, this is  exactly equivalent to what we get in our $m_q \neq 0, \to 0$ method, except that in our case the quasi-Goldstone pole is on the other side of the equation $Z_A (\Gamma_A)_{\mu}=...$,  see the r.h.s. of our eqn. (\ref{Award4}), second term, in $\Gamma_5$. The sign should be naturally opposite, since the contribution is on the other side of the W-T identity in the $m_q \to 0$ method ; its contribution to the pseudoscalar term is:
\begin{eqnarray}
 \frac { \langle\pi\vert \bar\psi \gamma_5 \psi \vert 0\rangle} {q^2-m_{\pi}^2}\Gamma_{\pi} (p,q);
\end{eqnarray}

\noindent and the contribution to eqn. (\ref{Award4} is then) :
\begin{eqnarray}
lim_{m_{\pi}^2 \to 0}~f_{\pi} \frac {\partial} {\partial q^{\mu}} \frac {m_{\pi}^2} {q^2-m_{\pi}^2}\Gamma_{\pi} (p,q)|_{q^{\mu}=0}=-f_{\pi} \frac {\partial} {\partial q^{\mu}}\Gamma_{\pi} (p,q)|_{q^{\mu}=0}
\end{eqnarray}
using  $f_{\pi} m_{\pi}^2=2 Z_A \rho \langle\pi\vert \bar\psi \gamma_5 \psi \vert 0\rangle$.
It is the same as in  eqn. (\ref{goldstone}), except that the sign is duely opposite. 

\vskip 0.2cm

Now,  the effect is present at any quark mass but going to the chiral limit is much easier  by our main method.

\end{document}